\documentclass[sigconf]{acmart}
\usepackage{booktabs} 
\usepackage{epstopdf}
\usepackage{epsfig}
\usepackage{chngpage}
\usepackage{float}
\usepackage{amsmath}
\usepackage{graphicx}
\usepackage{array}
\usepackage{epstopdf}
\usepackage{algorithm}
\usepackage{algorithmicx}
\usepackage{algpseudocode}
\usepackage{threeparttable}
\usepackage{subfigure}
\usepackage{multirow}
\usepackage{longtable}

\AtBeginDocument{%
  \providecommand\BibTeX{{%
    \normalfont B\kern-0.5em{\scshape i\kern-0.25em b}\kern-0.8em\TeX}}}

\copyrightyear{2020}
\acmYear{2020}
\setcopyright{acmcopyright}\acmConference[SIGIR '20]{Proceedings of the 43rd
International ACM SIGIR Conference on Research and Development in Information
Retrieval}{July 25--30, 2020}{Virtual Event, China}
\acmBooktitle{Proceedings of the 43rd International ACM SIGIR Conference on
Research and Development in Information Retrieval (SIGIR '20), July 25--30, 2020,
Virtual Event, China}
\acmPrice{15.00}
\acmDOI{10.1145/3397271.3401405}
\acmISBN{978-1-4503-8016-4/20/07}

\begin{document}
\fancyhead{}

\title{Web of Scholars: A Scholar Knowledge Graph}

\author{Jiaying Liu, Jing Ren, Wenqing Zheng}
\affiliation{
   \institution{School of Software \\Dalian University of Technology, China}
}
\email{{jiaying_liu,ch.yum}@outlook.com}

\author{Lianhua Chi}
\affiliation{
   \institution{School of Engineering and Mathematical Sciences \\La Trobe University, Australia}
}
\email{L.Chi@latrobe.edu.au}

\author{Ivan Lee}
\affiliation{
   \institution{UniSA STEM \\University of South Australia, Australia}
}
\email{Ivan.Lee@unisa.edu.au}

\author{Feng Xia}
\affiliation{
   \institution{School of Science, Engineering and IT \\Federation University Australia, Australia
}
}
\email{f.xia@ieee.org}
\renewcommand{\shortauthors}{J. Liu et al.}

\begin{abstract}
In this work, we demonstrate a novel system, namely Web of Scholars, which integrates state-of-the-art mining techniques to search, mine, and visualize complex networks behind scholars in the field of Computer Science. Relying on the knowledge graph, it provides services for fast, accurate, and intelligent semantic querying as well as powerful recommendations. In addition, in order to realize information sharing, it provides open API to be served as the underlying architecture for advanced functions. Web of Scholars takes advantage of knowledge graph, which means that it will be able to access more knowledge if more search exist. It can be served as a useful and interoperable tool for scholars to conduct in-depth analysis within Science of Science.
\end{abstract}

%
%



%

\maketitle

\section{Introduction}
In spite of the strong focus on scholarly information mining, retrieval, and utilization in the field of Science of Science, researchers still face various challenges in accessing to the accurate information with respect to state-of-the-art techniques. For instance, with the rapid growth of scholarly entities, it becomes more and more difficult to obtain information fast and accurately in large-scale networks. Meanwhile, redundant information on the Web also makes users be overwhelmed by the information, while many unstructured or semi-structured data are not put to great use. Due to the network complexity, how to efficiently mine implicit relationships among scholars is also a critical issue. Therefore, it is urgent to concentrate on managing the large amount of data and mining the implicit relationships among scholars to facilitate the comprehension of scholars.

To give a deeper understanding of academic social networks, some systems have already been developed to provide searching and mining services. For example, AMiner aims to extract information from the heterogeneous networks in the field of Computer Science. SCHOLAT and Social Scholar are Chinese websites to provide services of academic information management and literature retrieval. Unfortunately, most of them mainly focus on literature collection and simple explicit relationships exhibition (e.g., collaboration and citation relations), which makes it difficult to provide on-target services that can profile scholars based on users' requirements. The key challenge of providing such service is that scholarly networks include not only simple explicit but also complex implicit relationships (i.e., advisor-advisee relationship hidden in the collaboration network). It is difficult to utilize multiple scholarly data to construct multidimensional profiling for scholars.
\begin{figure}[t]
\centering
\includegraphics[width=0.39\textwidth]{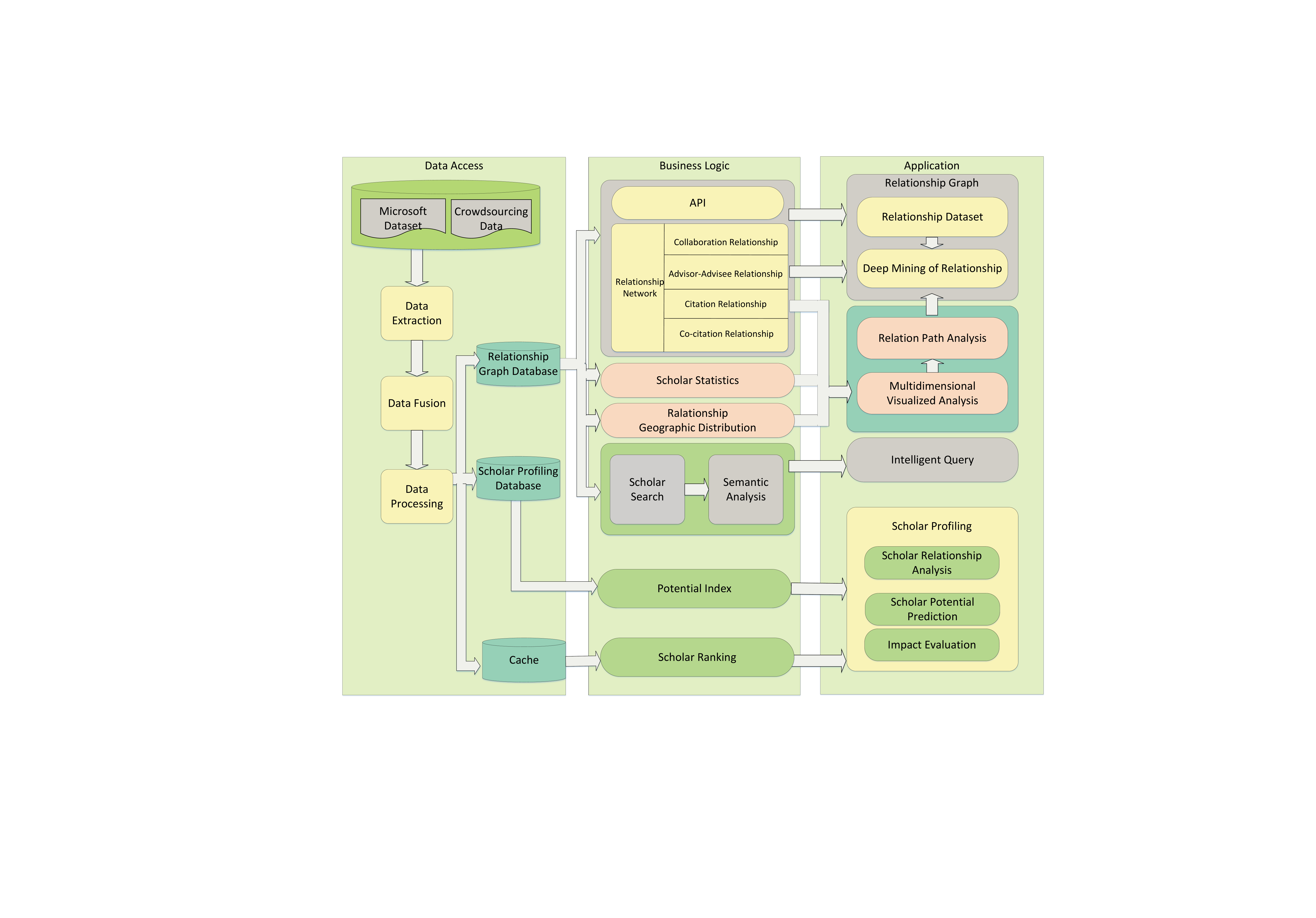} 
\caption{System Architecture of Web of Scholars.}
\label{figure:2}
\end{figure}

To fill these gaps, we present a knowledge graph-based system, namely Web of Scholars, to extract, integrate and profile relationships especially implicit relationships hidden behind scholars in detail, which in turn can construct scholars' relationship knowledge graphs and conduct visual presentations as well as interactive analysis. Taking advantage of knowledge graph, which combines theories and methods of applied mathematics, graphics, information visualization, and information science with citation analysis and co-occurrence analysis to visualize the core structure of the discipline~\cite{lin2015learning}, this system can achieve fast, accurate, and intelligent semantic search of scholarly entities. Furthermore, we utilize techniques of data mining, information processing, social network analysis, and graphic rendering to reveal and analyze implicit relationships such as advisor-advisee relationships in heterogeneous academic networks. Thus, the system aims to form a comprehensive service platform which is mainly available to scientific research institutions, and provides personalized services such as advisor recommendation, advisee recommendation, and expert finding for different users including students, scholars, and institutions. At the same time, the system also provides services such as information release.

The system takes a three-tier framework, including \emph{Data Access} layer, \emph{Business Logic} layer, and \emph{Application} layer (Figure~\ref{figure:2}) to achieve the goal of high-cohesion and low-coupling. The main functions in Web of Scholars include: (1) Scholar profiling, (2) Relationship knowledge graph, (3) Semantic analysis, (4) Intelligent query, (5) Academic ranking, (6) Scholar evaluation, (7) In-depth relationship mining and analysis, and (8) Visualization analysis.

Moreover, we construct external open API to allow users to download the relationship dataset, compile in-depth relationship mining, and develop knowledge-based inference system. We hope that it could be used as a tool for other advanced functions. For example, users are expected to expand the established relationship graph, update relation models, and convert them into a more complex network. It also encourages users to evaluate scholars, analyze and predict relationships among them to give a more accurate profiling of scholars, with the goal of constructing a complete academic network from the perspective of scholars. Other applications include: recommendations (e.g., reviewers, advisors, collaborators, team members), funding allocation, and in-depth scholarly social network analysis. The main contributions of this work are as follows:

\textbf{New Knowledge.} We present Web of Scholars as a novel system that tailors our generic methods to efficiently search, rank, and mine scholars as well as their various relationships from large heterogeneous academic networks, and exhibits visualization tools to present them. It also provides various academic applications such as advisor recommendation over a large academic relationship knowledge graph.

\textbf{Wealthy Information.} The system collects more than 1.7 million scholars, 1.5 million publications, and 7 different types total to over 433 million relationships among scholars. It attempts to clarify the complex academic network and storage the large knowledge graph in the graphic database.

\textbf{Tectonic Inheritance.} Web of Scholars is not only a search engine but also a bridge that can be served as the underlying architecture for other advanced functions such as reviewer recommendation, advisor choosing, team members recommendation, and funding allocation. It provides open API and allows integration into other environments for information sharing.

\section{System Overview}
\begin{figure}[t]
\centering
\includegraphics[width=0.39\textwidth]{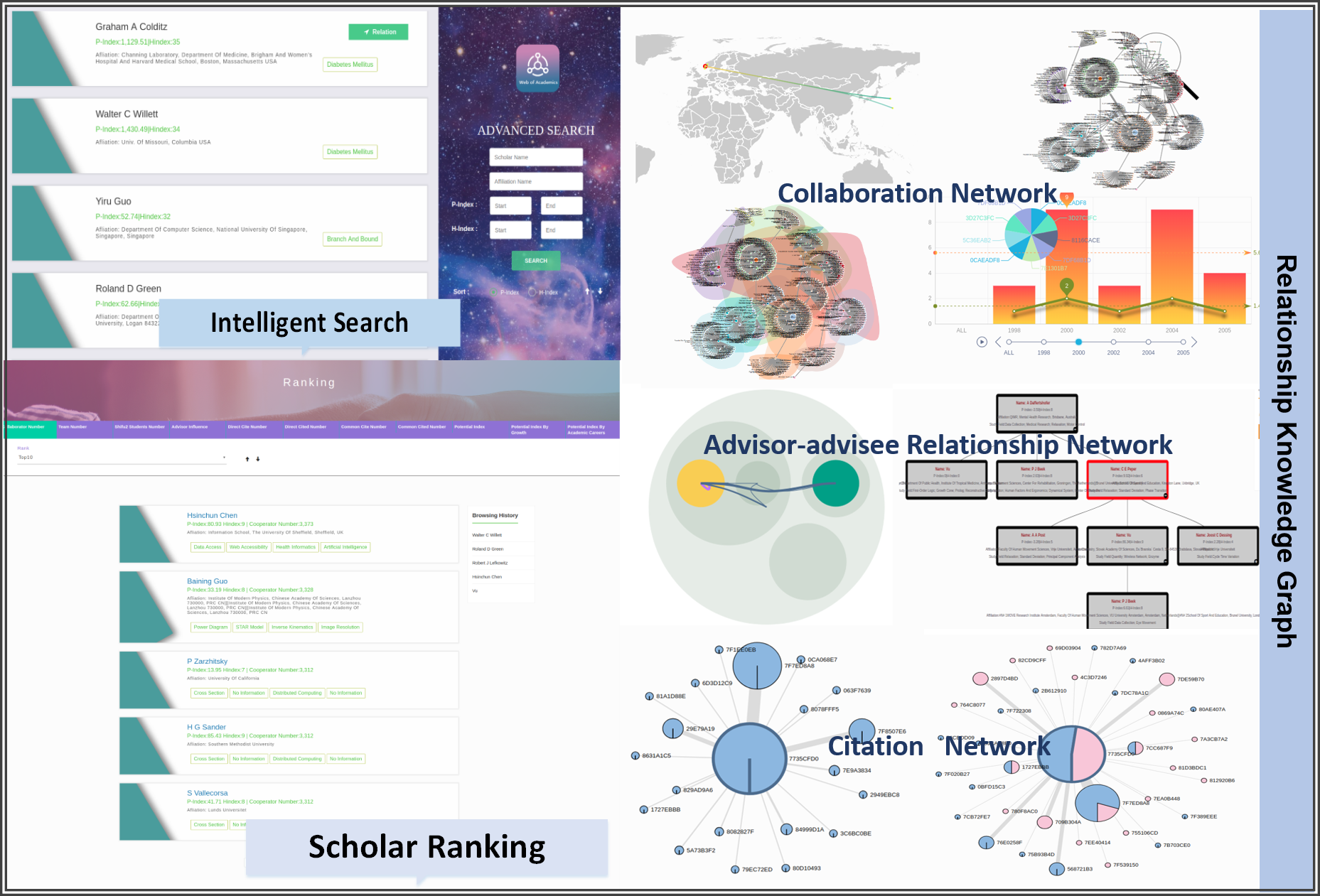} 
\caption{System Overview of Web of Scholars.}
\label{figure:1}
\end{figure}

Figure~\ref{figure:1} presents some of the main components of Web of Scholars:  1) The \emph{Profile Search} formulates queries of scholars including simple queries and intelligent queries. 2) The \emph{Academic Rank} is responsible for retrieving scholars in a descending order from the database under different categories. 3) Finally, the \emph{Relationship Knowledge Graph} is used to present scholars' various ego-networks, which aims to profile the scholar from the perspective of different relationships.

\subsection{Relationship Knowledge Graph}
\label{sec2}
The goal of Relationship Knowledge Graph is to profile scholars from the perspective of their relationships. It is comprised of several components, including the collaboration network, advisor-advisee relationships, and citation relationships.

\textbf{Collaboration Relationship.}
The co-author relationship includes three types of networks: ego-center collaboration network, the geographic distribution of collaborators, and changes of collaborator counts over time. We first extract all co-authors of the scholar $i$ from his/her publication metadata, then compute the times of collaborator relationship between them which can be represented by line thickness as shown in Figure~\ref{figure:1}.

To find scholars' geographic locations, we use Google Maps API to calculate the latitude and longitude of scholar's institution. Relying on d3.js, we facilitate the visualization of the collaborators in terms of geographic information. Beyond that, this part also displays the number of collaborator inactivity from 1980 to 2017, which is a good way to indicate the collaboration frequency.

\textbf{Advisor-advisee Relationship.}
The advisor-advisee relationship network aims to show the academic genealogical information of scholars. To attain the objective, it must take into account the following problems: \emph{given a scholar $i$, and his/her collaborator $j$, whether $j$ is $i$'s advisor?}

To address this problem, the system implements a network representation learning model, namely Shifu2~\cite{liu2019shifu2}, to generate the advisor-advisee dataset. To be specific, the process of dataset generation can be described as: 1) Crawl ground truth advisor-advisee pairs from the PhDTree project. 2) Match these pairs with scholars from Microsoft Academic Graph (MAG). 3) Differentiate scholars by academic age and calculate input features based on the publication information in MAG. 4) Use node attributes and edge attributes  to construct the node representations and the edge representations. Then put them into the classifier and optimize the model according to the precision. 5) Apply the model to the whole MAG dataset to gain a large-scale advisor-advisee pair dataset.

Furthermore, this part provides advisor recommendation service for students. It includes the processes of student preference collection, feature vector matching, recommendation results generation. Students need to fill in the characteristic of the advisors they want to find. The system will filter appropriate advisors in the database through feature matching, and finally show the recommendation results. In addition, the system will display the interface of details pages relationship networks of the recommended advisors, as well as the recommendation reason. Thus, students can understand the details of recommendation reason, so as to increase the trust and reliability of the results.

\textbf{Citation Relationship.}
For each scholar, we describe the citation network and the co-citation network which are different from traditional citation networks for papers. It is used to unveil scientific collaboration patterns and mine the implicit relationships among scholars. For example, in the citation network, we distinguish the identity (e.g., advisors, advisees, co-authors, or collaborators) of the referee with color. We also use different colors and sizes to represent the importance of nodes in scholars' co-citation networks.

\subsection{Academic Rank}
\label{ranking}
In the ranking model, we define different measures to evaluate scholars' achievement, including ``Number of Collaborators", ``Number of Advisees", ``Number of Team Members", ``Advisor Influence", ``Times of Citations", and ``Potential Index". For each measure, the system outputs a descending ranking list in the domain of Computer Science. We store the ranking in the Redis cluster. This cached approach can prevent needless round trips to the database, thus ensuring the quickness and timeliness when visiting rankings.

\section{Demonstration}
\subsection{Implementation}
As mentioned above, the system consists of three layers:

\emph{Data Access} layer leads to implementation operations of relationships in the graphical database and data in the NoSQL database. In this system, we mainly use TITAN, which is a scalable graph database optimized for storing and querying graphs. The graphical database employs HBase to achieve data storage and retrieval.

\emph{Business Logic} layer uses Spring MVC Framework to make it interact with the front-end data. To simplify the development process, Spring container is used to manage Java Bean through the container injection method.

\emph{Application} layer is located at outermost, which is closest to users and could be used to receive the input data and present it. It can also provide users with an interactive operation. The front-end exhibits an elastic template engine, FreeMarker to generate output. The visualization of various networks is implemented using d3.js toolkit. For rendering and landscaping them, we use Bootstrap.

Moreover, to ensure the system can normally run with the server temporarily breakdown, a distributed environment, \emph{Hadoop + Hbase + Zookeeper} with three server machines on Linux operating system is designed.

\subsection{Data}
For implementing Web of Scholars, we rely on the popular digital library, Microsoft Academic Graph (MAG), which contains 171,233,592 publications, 209,508,429 authors, and 196,025 research fields. Based on the research field, we extract all publications in the field of Computer Science, and identify all scholars in this field to generate relationships mentioned above in Section~\ref{sec2}. Finally, by processing the publication metadata, Web of Scholars collects more than 1.7 million researchers, 1.5 million papers, 20 million collaboration relationships, 1.4 million collaborative teamworks, 1 million advisor-advisee relationships, 14 million citation relationships, and 300 million co-citation relationships. All of them are stored in the graphic database in terms of knowledge graph.

\subsection{Interface}
Web of Scholars provides two kinds of interfaces: \textbf{User Search Interface} and \textbf{Open Interface}. User Search Interface aims to fulfill user expectations of interacting with the system. The query input panel implements a fuzzy matching technique provided by HBase, FuzzyRowFilter, where any token of searching names (synonyms) could be auto-completed. It can also achieve the goal of intelligent query. For example, typing ``Bob's advisor" retrieves his advisor as a suggestion. We also provide most of Web of Scholars' API to users freely. Users can share copies and redistribute the data in any format for free to build advanced functions.

\subsection{Screen Shots}
\begin{figure*}[t]
\centering
\includegraphics[width=0.68\textwidth]{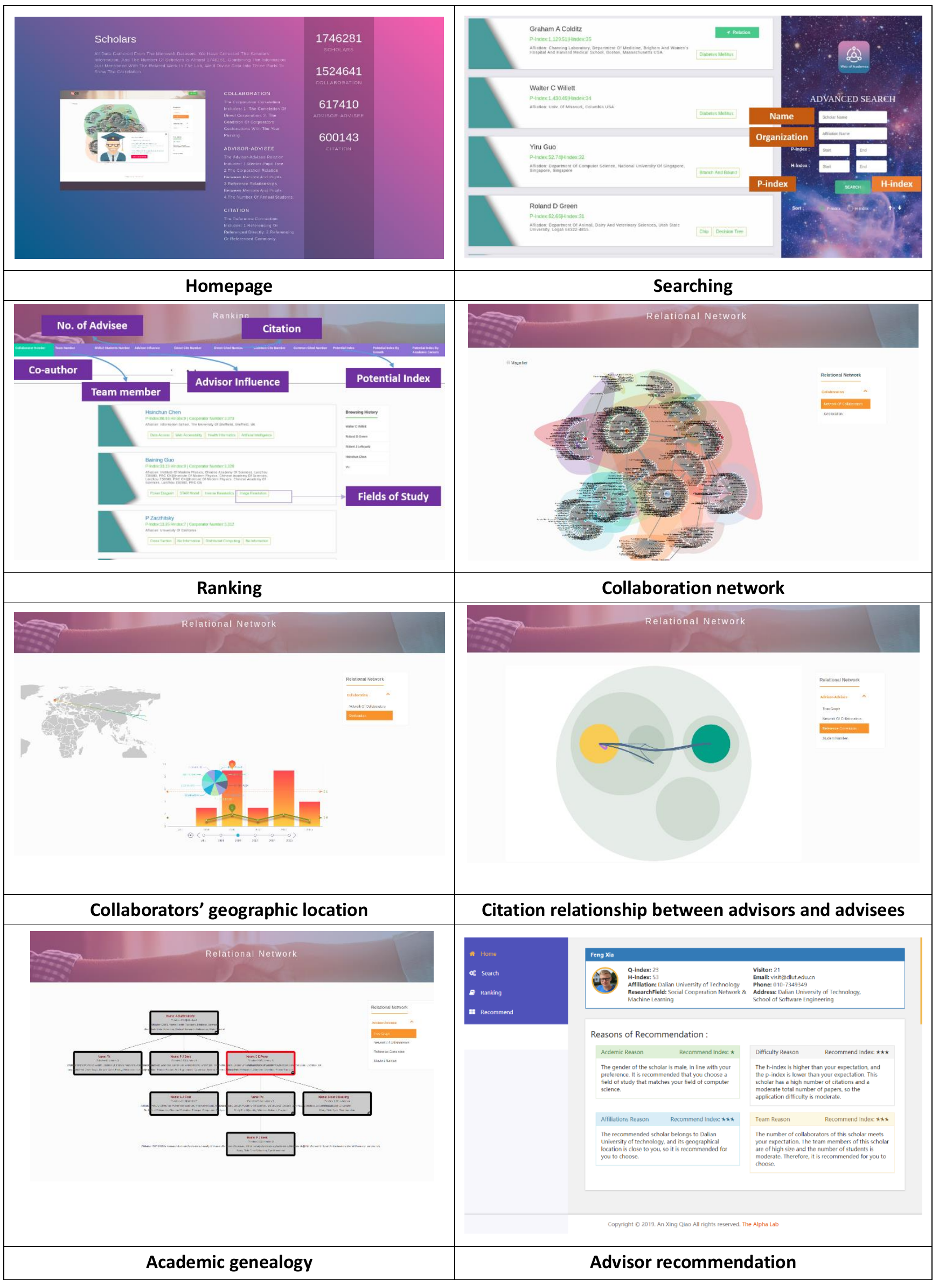} 
\caption{Some Screen Shots of Web of Scholars.}
\label{figure:3}
\end{figure*}

Figure~\ref{figure:3} shows some screen shots of the system. The URL\footnote{http://thealphalab.org/resources.html} presents a 5-minute video of Web of Scholars.

\section*{Acknowledgement}
This work is partially supported by National Natural Science Foundation of China under Grant No. 61872054 and the Fundamental Research Funds for the Central Universities (DUT19LAB23). The authors would like to thank Huijie Zhang and Wenjie Kang for help with system design.
%
\bibliographystyle{ACM-Reference-Format}
\bibliography{samplebibliography}

\end{document}